\documentstyle[12pt]{article}

\def\HS{Haldane--Shastry}
\def\ket[#1]{\:\vert\,{#1}\,\rangle}
\def\notin{{\in}\kern-0.4em{/}\kern+0.05em}
\begin{document}

\begin{titlepage}

\def\thefootnote{\fnsymbol{footnote}}
%%
%% date/preprint no(s). etc:
%%
\begin{flushleft}
ITP-UH-07/94 \hfill April~1994\\
cond-mat/9405038
\end{flushleft}

\vspace*{\fill}

\begin{center}
{\Large NEW FAMILY OF SOLVABLE 1D HEISENBERG MODELS} \\
\vfill
%%
%% Author(s):
%%
{\sc Holger Frahm\vspace{0.5em}}\footnote{%
e-mail: {\tt frahm@itp.uni-hannover.de}}\\
{\sl Institut f\"ur Theoretische Physik, Universit\"at Hannover,
D-30167~Hannover, Germany\vspace{1.5em}}\\
{\sc Vladimir I.\ Inozemtsev\vspace{0.5em}}\footnote{%
e-mail: {\tt inozv@theor.jinrc.dubna.su}}\\
%%
%% Address:
%%
{\sl Laboratory of Theoretical Physics, JINR, 141980 Dubna, Russia}\\
\vfill
ABSTRACT
\end{center}
\setlength{\baselineskip}{13pt}
\begin{quote}
Starting from a Calogero--Sutherland model with hyperbolic interaction
confined by an external field with Morse potential we construct a
Heisenberg spin chain with exchange interaction $\propto 1/\sinh^2 x$ on a
lattice given in terms of the zeroes of Laguerre polynomials. Varying the
strength of the Morse potential the Haldane--Shastry and harmonic spin
chains are reproduced. The spectrum of the models in this class is found to
be that of a classical one-dimensional Ising chain with nonuniform nearest
neighbour coupling in a nonuniform magnetic field which allows to study the
thermodynamics in the limit of infinite chains.
\end{quote}

\vfill
\noindent PACS-numbers:
75.10.Jm~\  %Heisenberg and other quantized localized spin models
05.30.-d~\  %Quantum statistical mechanics
%%03.65.Ca~\  %Quantum theory; quantum mechanics
05.50.+q~\  %Lattice theory and statistics; Ising problems
%% 05.70.Jk~\  %Thermodynamics

\vfill
% Short title:
% %            12345678901234567890123456789012345678901234567890
\setcounter{footnote}{0}
\end{titlepage}

Studies of Heisenberg spin chains with exchange coupling proportional to
certain functions decaying like the inverse square of the distance have
unveiled surprisingly simple properties which allow for a detailed study of
these many-body systems. Among of these properties are exact wave functions
of Jastrow product form and very simple spectra.  Most notably, the
Haldane--Shastry model \cite{hald:88,shas:88} with trigonometric ($\propto
1/\sin^2x$) exchange admits an interpretation as the generalization of the
concept of an ``ideal gas'' to the case of fractional statistics
\cite{hald:91,bpd:93,hald:94}.

All of these models can be related to Calogero--Sutherland models of
particles with an internal degree of freedom (spin) using the exchange
operator formalism \cite{poly:92}. Within this approach, the spin chains
are obtained by ``freezing out'' the kinematic degrees of freedom in these
models thereby giving a pure exchange model of particles interacting with
inverse square potential on a lattice given by the static equilibrium
positions of the Calogero model. Within this formalism it has been possible
to construct a family of operators commuting with the Hamiltonian and among
themselves for the Haldane--Shastry model \cite{fomi:93} and a model with
rational exchange related to the $1/x^2$ Calogero model confined by a
harmonic potential well \cite{poly:93,frah:93,poly:94} (due to its spectrum
this model will be referenced as the ``harmonic spin chain'' in the
following).  Although it has not been possible so far to obtain the
Hamiltonian from this construction directly this is to be considered as the
proof of integrability of these models.

In the limit of infinite chains the above considerations have been applied
to the hyperbolic chain with exchange $\propto 1/\sinh^2x$ on a
translationally invariant lattice. A generalization of this case to {\em
finite nonuniform} lattices which reduces to the \HS\ and the harmonic spin
chain in certain limits is presented in this letter. We start with the
Hamiltonian of a $N$-particle Calogero--Sutherland system of particles
interacting with an external field with Morse potential given by
\begin{equation}
   {\cal H}_{CS} = \sum_{j}^{N}\left[{{p_{j}^{2}}\over2}
                       +2\tau^2(e^{4x_{j}}-2e^{2x_{j}})\right]
        +\sum_{j<k}^{N}{1\over{\sinh^2(x_{j}-x_{k})}}\ .
   \label{calo_hamil}
\end{equation}
The classical equilibrium positions of the particles (that will later
define the lattice on which the spins are placed) are given in terms of the
set of nonlinear algebraic equations
\begin{equation}
   -\sum_{k\neq{j}}^{N}{{z_{k}(z_{j}+z_{k})}\over{(z_{j}-z_{k})^3}}
   +\tau^2(z_{j}-1)=0\ (j=1,\ldots,N)\ ,
   \label{the_lattice}
\end{equation}
where we have introduced $z_j=e^{2x_j}$. A general solution to
(\ref{the_lattice}) is unknown.  However, following the observation by
Calogero \cite{calo:77} for $1/x^2$ case we assume that the $\{z_j\}$ are
the roots of some polynomial $p_N(z) = \prod_{j=1}^N \left(z-z_j\right)$
obeying the second order differential equation \cite{inoz:89a}
\begin{equation}
   Ap_{N}''+Bp_{N}'+\lambda_{N}p_{N}=0\ ,
   \label{diffp_1}
\end{equation}
where $A$ and $B$ are polynomials in $z$ and $\lambda_N$ is some constant.
Choosing $\lambda_N=2N\tau$ we find that (\ref{diffp_1}) reads
\begin{equation}
   y{{d^2p_{N}(y)}\over{dy^2}}+
   \left(-y + 1+\Gamma\right){{dp_{N}(y)}\over{dy}}+Np_{N}(y)=0\ ,
   \qquad y=2\tau z
   \label{diffp}
\end{equation}
with $\Gamma=2(\tau-N)+1$ from which we identify $p_N$ as the Laguerre
polynomial $L_N^{(\Gamma)}(2\tau z)$. Let us note the following
properties of $L_{N}^{(\Gamma)}$:
\begin{enumerate}
\item
For $\Gamma>-1$ all roots of $L_{N}^{(\Gamma)}$ are real positive
numbers.
\item
$L_{N}^{(\Gamma)}$ has multiple zeroes if and only if $\Gamma=-n$,
$n=2,3,\ldots,N$. At these values of $\Gamma$ one can write
\[
   L_N^{(-n)}(y) = {(N-n)!\over N!} (-1)^{(n)}
    y^{n} L_{N-n}^{(n)}(y)\ .
\]
\item
As $\Gamma = -n+\varepsilon$, $\varepsilon\to 0$ the $n$ roots of
$L_{N}^{(\Gamma)}$ approaching 0 have the following asymptotic behaviour:
\begin{equation}
   z_{j}\sim\hbox{const}\vert\varepsilon\vert^{{1\over{n}}}
      \exp\left({{2\pi{i}j}\over{n}}\right) ,\quad j=1,\ldots,n\ .
   \label{roots_u}
\end{equation}
\end{enumerate}
Introducing variables $z_j=1+\tau^{-{1\over2}}\zeta_j$ and taking the limit
of large $\tau$ the rational limit of the Calogero model related to the
harmonic spin chain is obtained. In this case the lattice points $\zeta_j$
are the roots of the Hermite polynomial $H_N(\zeta)$.  For the special
values of $\Gamma$ being a negative integer $\ge -N$ the lattice separates
into two parts not coupled by the interaction term in the Hamiltonian. Upon
rescaling one of them is given by $n$-th roots of unity (\ref{roots_u}),
which results into a lattice of equally spaced sites with periodic boundary
conditions. The corresponding spin chain is the \HS\ model with interaction
$\propto 1/\sin^2x$. The other part corresponds to the hyperbolic model
with $\Gamma$ replaced by $-\Gamma$. Finally, as $\Gamma\to -N$ the model
reduces to the trigonometric model.

Following Fowler and Minahan \cite{fomi:93} we now consider the a model
where $N$ bosonic particles are sitting on different points $z_j$
satisfying (\ref{the_lattice}) and allow the exchange of particle positions
as the only dynamical process. Denoting the corresponding Hermitian
exchange operator for particles $i$ and $j$ by ${\cal M}_{ij}$ we choose
the Hamiltonian of this system to be
\begin{equation}
   {\cal H}_{ex} = \sum_{j<k}^N h_{jk} {\cal M}_{jk}\ ,\qquad
   h_{jk} = {z_j z_k \over (z_j - z_k)^2}\ .
   \label{hamil_x}
\end{equation}
We now construct operators
\begin{equation}
   h_{j}={1\over4}
   (\hat \pi_{j}+\omega_{j})(\hat \pi_{j}-\omega_{j})
            -\tau^2 (z_j-1)^2 + \hbox{const}\ \omega_{j}
   \label{def:hj}
\end{equation}
where
\[
   \hat \pi_{j}=\sum_{k\neq{j}}^{N}{{z_{j}+z_{k}}
                      \over{z_{j}-z_{k}}}{\cal M}_{jk}\ , \qquad
   \omega_j = \sum_{k\ne j}^N {\cal M}_{jk}\ .
\]
At some value of the constant in (\ref{def:hj}) these operators can be
factorized:
\begin{equation}
   h_{j}=a_{j}^{+}a_{j}^{-},\quad
   a_{j}^{\pm}={1\over2}\left(\hat \pi_{j}\pm\omega_{j}\right)
           \pm\delta\tau(z_{j}-1)\ ,\quad \delta^2=1\ .
\end{equation}
Commutators of two operators from this set are a bit more complicated
than in the case of harmonic chain,
\[
   [h_{j},h_{k}]=2\delta\tau(h_{j}{\cal M}_{jk}-{\cal M}_{jk}h_{j})
   +\omega_{j}{\cal M}_{jk}h_{j}-h_{j}{\cal M}_{jk}\omega_{j}\ .
\]
Using the properties
\[
   {\cal M}_{ij} z_i = z_j {\cal M}_{ij}\ ,\quad
   {\cal M}_{ij} z_k = z_k {\cal M}_{ij}\quad \hbox{for~} i\ne k\ne j
\]
\[
   {\cal M}_{ijk} = {\cal M}_{ik}{\cal M}_{ij}
                  = {\cal M}_{ij}{\cal M}_{jk}
                  = {\cal M}_{jk}{\cal M}_{ik}
\]
of the exchange operators ${\cal M}_{ij}$ one finds that the commutators
$\left[ h_j, {\cal H}_{ex} \right] = 0$ for an {\em arbitrary} set of
solutions to (\ref{the_lattice}) and hence the operators
\begin{equation}
   {\cal I}_m = \sum_{j=1}^N h_j^m
\end{equation}
all commute with the Hamiltonian (\ref{hamil_x}). For the \HS\ model
\cite{fomi:93} and the harmonic spin chain \cite{poly:93} it can be shown
that operators similar to these ${\cal I}_m$ also commute among themselves.
We have not been able to prove this in the general case considered
here. After tedious calculations we were only able to show that
$\left[{\cal I}_n, {\cal I}_m\right]=0$ for $1\le n,m \le 3$.

As usual the operators ${\cal M}_{jk}$ can be related to exchange operators
acting not on the particle positions but on their internal degree of
freedom as they are defined in the space of bosonic, i.e.\ totally
symmetric wavefunctions. This leads us to consider the Hamiltonian
\begin{equation}
   {\cal H} = -\sum_{j<k}^{N} h_{jk}
       {{\vec \sigma_{j}\cdot\vec\sigma_{k}-1}\over2}
            = C_N - \sum_{j<k}^{N} h_{jk} {\cal P}_{jk}\ ,\qquad
   C_N = {1\over 24} N(N-1) \left(3\Gamma + 2N-1 \right)
   \label{spin_hamil}
\end{equation}
with the spin exchange operator ${\cal P}_{jk}=
{1\over2}\left(\vec\sigma_{j}\cdot\vec\sigma_{k} -1\right)$.  The
$\sigma^\alpha_j$ are Pauli matrices acting on the $j^{\hbox{th}}$ site of the
lattice. The same procedure gives a trivial exchange operator $\tilde{\cal
I}_1 = -{1\over 3} \sum {\cal P}_{ijk}$ for the spin analogue of ${\cal
I}_1$. The next operator in this sequence, $\tilde{\cal I}_2$, already
contains four-spin exchange terms with nonuniform exchange couplings.

To construct eigenstates of the Hamiltonian (\ref{spin_hamil}) we start
from the ferromagnetic vacuum
$\ket[0]=|\uparrow\uparrow\ldots\uparrow\rangle$ and consider states with
given magnetization
\begin{equation}
    |\psi^{(M)}\rangle = \sum_{j_{1}<...<j_{M}}^{N} \psi(j_{1}...j_{M})
         \prod_{s=1}^{M}\sigma_{s}^{-}\ket[0]\ .
\end{equation}
Using the fact that the $z_j$ are roots of the $N$-th Laguerre polynomial
we find that the eigenstates in the one-magnon sector ($M=1$) have
amplitudes
\begin{equation}
   \psi_{m}(j) \propto z_{j}^{m} {
         {L_{N-m-1}^{(\Gamma+2m)}(2\tau{z}_{j})}\over
         {L_{N-1}^{(\Gamma)}(2\tau{z}_{j})}}\
    ,\qquad m=0,\ldots,N-1\ .
   \label{states1}
\end{equation}
Their energies $E_{m}^{(1)}=\epsilon_m$ are given by the following quadratic
dispersion law:
\begin{equation}
   \epsilon_m = {m \over 2} \left( \Gamma + m \right)\ .
   \label{dispersion}
\end{equation}

Next we have studied the two-magnon sector ($M=2$). The amplitudes
$\psi(j_1,j_2)$ can be written as polynomials in $\left\{
z_{j}^{-1}\right\}$. No closed expression for these amplitudes has been
found. Nevertheless, the eigenproblem can be solved analytically and we
were able to find the complete set of $N(N-1)/2$ eigenvalues which can be
written as
\begin{equation}
   E_{m,n}^{(2)} = \epsilon_m + \epsilon_n\left(1-\delta_{m,n-1}\right)\ ,
     \qquad 0\le m < n \le N-1
\end{equation}
with the single magnon energies $\epsilon_m$ given by (\ref{dispersion}).

Finally we considered the $M$-magnon sector. Within the Ansatz
\begin{equation}
     \psi={{
     \prod_{\lambda>\mu}^{M}(z_{j_{\lambda}}-z_{j_{\mu}})^2
        F(z_{j_{1}}...z_{j_{M}})} \over
     {\prod_{\nu=1}^{M}z_{j_{\nu}}p'_{N}(z_{j_{\nu}})}}
\end{equation}
where $F$ is some symmetric polynomial in $\{z\}$, the corresponding part
of spectrum which comprises ${{(N-M+1)!}\over{M!(N-2M+1)!}}$ eigenvalues is
found to be additive
\begin{equation}
   E_{\{m_k\}}^{(M)} = \sum_{k=1}^M \epsilon_{m_k}
\end{equation}
with the dispersion (\ref{dispersion}) where the integers $m_k$ obey the
set of restrictions
\begin{equation}
   0 \le m_1 < m_2-1 < \ldots < m_{M} -1 \le N-2\ .
\end{equation}

While our solution of the $M$-particle sector is not complete we make the
following hypothesis concerning the spectrum of the class of models
(\ref{spin_hamil}):

All the eigenvalues of ${\cal H}$ can be written in compact form
\begin{equation}
    E_{n_{1}...n_{N}}=\sum_{k=1}^{N-1}\epsilon_{k}n_{k+1}(1-n_{k})\ ,
   \label{isingspec}
\end{equation}
where $\epsilon_{k}={1\over2} k\left(\Gamma+k\right)$ and $\{n_k\}=0,1$. As
a consequence, the operator ${\cal H}=-1/2\sum_{j<k}^{N}h_{jk}\vec
\sigma_{j}\vec\sigma_{k}$ is equivalent up to unitary transformation
to the Hamiltonian of classical 1D Ising chain in a non-uniform magnetic
field
\begin{eqnarray}
   {\cal H}_{I}&=&\sum_{k=1}^{N-1}\epsilon_{k}(\sigma_{k+1}-\sigma_{k}-
          \sigma_{k+1}\sigma_{k})
   \nonumber \\
   &=&\epsilon_{N-1}\sigma_{N}+
      \sum_{k=0}^{N-2}[\sigma_{k+1}(\epsilon_{k}-\epsilon_{k+1})
         -\sigma_{k+1}\sigma_{k+2}\varepsilon_{k+1}]
   \nonumber
\end{eqnarray}
with $\{\sigma_{k}\}=\pm1$.

We have checked this hypothesis numerically to give the correct spectrum
for lattices of lengths up to $N=12$. We have also confirmed it in the
limit $\Gamma\to\infty$ by establishing the analytic correspondence to the
effective Hamiltonian of the harmonic chain (Eq.~(20) in \cite{frah:93}).
Finally, it is consistent with the picture of lattice separation at
$-N\le\Gamma=-n<-1$ (where $\epsilon_{n}=0$) mentioned above and coincides
with the Ising Hamiltonian that follows for the \HS\ model in the
appropriate limit of the Hubbard model with $1/r$ hopping \cite{geru:92}.

The simple form of the spectrum (\ref{isingspec}) allows to compute the
free energy in the thermodynamic limit by using the transfer matrix
method. The partition function on a finite lattice can be written as
\begin{eqnarray}
   {\cal Z}_N = {1\over 2}
      \hbox{trace}\left\{
      \left( \begin{array}{cc} 1/w_1 & -1/w_1 \\ 1 & 1
             \end{array} \right)
      \prod_{k=1}^{N-2} \left[
      \left( \begin{array}{cc} 1+w_k & 0 \\ 0 & 1-w_k \end{array} \right)
      {\cal B}_k \right] \right. \times \quad && \nonumber \\
    \times \left.
      \left( \begin{array}{cc}
        ( 1+w_{N-1})^2 & ( 1+w_{N-1})^2 \\
        ( 1-w_{N-1})^2 & ( 1-w_{N-1})^2
	     \end{array} \right) \right\} &&
    \label{partf_n}
\end{eqnarray}
where
\begin{equation}
      {\cal B}_k = {1\over 2} \left( \begin{array}{cc}
          1 + {w_k\over w_{k+1}} & 1 - {w_k\over w_{k+1}} \\
          1 - {w_k\over w_{k+1}} & 1 + {w_k\over w_{k+1}}
	     \end{array} \right)\ ,
      \qquad
      w_k = e^{-{1\over 2}\beta \epsilon_k}\ .
\end{equation}
To perform the thermodynamic limit $N\to\infty$ we rescale the magnon
energies (\ref{dispersion}) with a factor $1/N^2$ and find the leading term
of (\ref{partf_n}) to be $Z \sim \exp(\beta E_0)\prod_{k=1}^{N-2}
\left(1+w_k\right)$ with some proper $E_0$ renormalizing the ground state
energy to zero. {}From this expression one obtains for the free energy per
site
\begin{equation}
   f = - {E_0\over N} - {1\over \beta}
      \int_0^1 dx \ln\left( 1+e^{-\beta \epsilon(x)/2} \right)\ .
\end{equation}
Using the quasiparticle dispersion $\epsilon(x) = x (\gamma + x)$ (the
renormalization of the magnon energies leads to $\gamma=\Gamma/N$ when
taking the thermodynamic limit of (\ref{dispersion})) this can be written
for $-1< \gamma <0$ as
\begin{equation}
   f = -{1\over \beta} \left(
    \int_0^{-\gamma} dx \ln\left(1+e^{ \beta \epsilon(x)/2} \right)
  + \int_{-\gamma}^1 dx \ln\left(1+e^{-\beta \epsilon(x)/2} \right)
    \right)
\end{equation}
Note that changing the sign of the quasiparticle dispersion or,
equivalently, the exchange in (\ref{spin_hamil}) from ferromagnetic to
antiferromagnetic changes the free energy by a temperature independent term
(due to the different ground state energy) only. This has been noticed in
the case of the Haldane--Shastry model $\gamma=-1$ before \cite{hald:91}
and is a consequence of the possibility to describe these models in terms
of an effective Ising model (\ref{isingspec}).

Since the total spin of the effective Ising model is proportional to the
$z$-component of the total spin in the original model one can also study
the effect of an external magnetic field. For $\Gamma>-1$ the largest
eigenvalue of (\ref{spin_hamil}) in the $M$-magnon sector is
\begin{equation}
   E_{max} = {1\over 6} M\left( 3(N-M)(N-M+\Gamma) + M^2-1\right)
   \qquad \hbox{for~} M\le {1\over 2}N
\end{equation}
from which we conclude that the ground state magnetization of the {\em
anti}ferromagnetic chain in a magnetic field $h$ is given by ${\cal M}(h) =
{1\over 4} \left(\sqrt{\gamma^2+8h} - \gamma\right)$ for $h\le
h_s={1\over2}\left( \gamma+1 \right)$.  Beyond $h_s$ the magnetization is
saturated at ${1\over2}$. For finite temperatures the transfer matrix
method yields
\begin{equation}
   {\cal M}(\beta,h) = {1\over 2} \int_0^1 dx
     {\sinh\beta h/2 \over \sqrt{\exp(-\beta\epsilon(x)) +
                                 \sinh^2 \beta h/2}}\ .
\end{equation}

In this letter we have constructed a new family of solvable Heisenberg spin
chains. As in the previously known \HS\ and harmonic models their spectra
can be given in terms of an effective Ising Hamiltonian. This mapping
allows to study the thermodynamics of this models in detail.
Generalization to other exchange type models such as interacting $SU(N)$
spins or electrons subject to a supersymmetric $t$--$J$ or Hubbard type
Hamiltonian are easily constructed by considering different representations
of the permutation operators in (\ref{spin_hamil}). At the same time, a
number of interesting questions, especially regarding the construction of a
complete set of conserved quantities, remains open.

This work has been partially supported by the Deutsche
Forschungsgemeinschaft under grant no.\ Fr~737/2--1.  One of the authors
(V.I.) has received partial support within the framework of the
Heisenberg--Landau program.

\newpage

%%\bibliography{base}
%%\bibliographystyle{mybib}

\end{document}